# Extending the Pnictide Chemical Space for Photovoltaics

Avaneesh Balasubramanian[1,2] and Gopalakrishnan Sai Gautam[2]*

[1] Indian Institute of Science Education and Research, Pune 411008, India

[2] Department of Materials Engineering, Indian Institute of Science, Bengaluru 560012, India

* E-mail: saigautamg@iisc.ac.in

## Abstract

Developing novel and efficient materials that are beyond silicon for photovoltaic (PV) applications is required to meet the upcoming energy needs of our societies. To identify novel candidate materials that can act as PVs, we use first principles calculations to perform a systematic screening of the pnictide chemical space (i.e., nitrides and phosphides). Specifically, we explore three different ternary and quaternary pnictide classes, namely, $ABCX_2$, $BB'B''X_2$, and $A_4BX_2$ (A = Li, Na, or K; B, B', B" = Ca, Sr, Mg, or Zn; C = Al, Ga, or In; X = N or P), leading to a set of 104 possible pnictide compositions. Based on our evaluations of ground state structures, 0 K thermodynamic stabilities, electronic structures, carrier effective masses, dynamic stabilities, and intrinsic point defect formation energies, we arrive at three promising candidates, namely, $NaCaInN_2$, $NaSrInN_2$, and $K_4ZnP_2$. Notably, all the identified candidates are thermodynamically (meta)stable, exhibit direct (or nearest direct) band gaps that are optimal for PV applications, and are resistant to forming several types of point defects. We hope that our first principles driven workflow and the identified candidates will advance the development of novel PV materials and reinvigorate interest in the exploration of pnictides.

## 1 Introduction

The drastic increase in world population demands energy and resources to a larger extent than ever before. The conventional energy sources that have been catering to society's needs are non-renewable fossil fuels, which, on combustion, release greenhouse gases known to cause extreme weather events and climate change [1]. On the other hand, converting solar energy into electricity by photovoltaic (PV) devices (i.e., solar cells) significantly reduces the dependence on non-renewable energy sources [2]. The state-of-the-art silicon (Si) solar cells typically contain thick sheets (200-500 μm) of Si for sufficient light absorption due to its intrinsic indirect band gap, thus increasing its manufacturing, installation, and maintenance costs [3, 4]. Therefore, the exploration of other stable materials that are earth abundant and display a direct band gap within the Shockley-Queisser limit of single junction PVs that maximizes solar cell efficiency (~1.1 to 1.5 eV), is both important and relevant [5].

Besides an optimal band gap, the semiconductors competing with Si as possible PVs must be made of non-toxic and earth-abundant elements, be stable at the temperature at which they are synthesized and operational, and resist forming intrinsic point defects, since defects can act as (deep) trap states for the charge carriers in the device (thus modifying electronic properties) and can also modify the structure of the material. High charge carrier mobility (i.e., smaller effective masses [6]) and lifetimes are important as well for the PV performance of the material. Among beyond-Si materials, pnictides, which are compounds with group 15 elements in their -3 oxidation state, such as GaAs [7], and (B, Ga, In)Sb [8-10], have been well-studied and optimized for PV applications. However, the presence of $As^{3-}$ or $Sb^{3-}$ make these materials toxic during disposal and impractical for widespread adoption. Previous research on safer binary and ternary pnictides, containing $N^{3-}$ and $P^{3-}$, has highlighted their potential as candidate PV materials [4, 11]. Nitrides, in particular, are known for their high thermal stability and ideal band gaps due to a better matching of the electronegativity of nitrogen with group 13 elements like Ga [12]. Notably, our previous *ab initio* study [4] established that among binary $A_3B_2$, ternary $AA'B_2$, and quaternary $AA'A''B_2$ (where A,A′, and A″ = Sr, Zn, and/or Ca and B = N or P), $SrZn_2N_2$, $SrZn_2P_2$, and $CaZn_2N_2$ can be strong candidates for beyond-Si PV applications. However, the chemical space used in [4] is limited to pnictides containing select group 2 cations only, besides Zn, indicating a need to further expand the search space for identification of other novel PV candidates.

Previous work done on specific pnictide compounds, such as $K_4CdP_2$ and $K_4ZnP_2$, have shown that the electronic structure of such compounds could make them useful for PV technologies [13]. More importantly, such studies have indicated the promise of pnictide compositions consisting of non-group-2 elements, such as alkali metals. However, no comprehensive computational or experimental study has screened through the broad pnictide chemical space, especially ternaries and quaternaries. Even for a select set of pnictide compounds with known crystal structures in chemical databases, thermodynamic and electronic data relevant for PV applications is missing, which is the motivation of this study.

Here, we perform a high-throughput screening of 104 pnictide compositions, comprising $ABCX_2$, $BB'B''X_2$ and $A_4BX_2$ classes (A = Li, Na, or K; B′, B″ = Ca, Sr, Mg, or Zn; C = Al, Ga, or In; X = N, P) using density functional theory (DFT [14, 15]) based calculations. Specifically, we identify the stable (ground state) structure and construct the 0 K convex hull for determining the thermodynamic stability [16] for each composition. Subsequently, we filter the 104 compounds sequentially based on their electronic band gap from density of states (DOS) calculations, hybrid-functional-based band structure evaluations indicating the nature of the band gap and the associated carrier effective masses, phonon properties quantifying dynamic stability, and intrinsic point defect formation energies. Unlike our previous work [4], we consider Mg-compounds for our PV screening to ensure that our chemical space remains wide. Based on our calculated data and screening, we identify three promising candidates for PV applications, namely, $K_4ZnP_2$ of the $A_4BX_2$ class, and $NaCaInN_2$ and $NaSrInN_2$ of the $ABCX_2$ class, based on their thermodynamical and dynamical stability, optimal band gaps (hybrid-functional-calculated band gap of 1.1-1.73 eV), and resistance to the formation of point defects. Note that although $Na_4MgP_2$ and $Na_4ZnP_2$ display an optimal band gap, $Na_4MgP_2$ is not resistant to point defect formation (defect formation energy $< 1$ eV) and $Na_4ZnP_2$ shows dynamic instabilities, highlighting the need to incorporate multiple properties in the screening process to precisely identify robust candidates. We hope our study encourages further work on the design of high-efficiency PV devices within the pnictide chemical space.

## 2 Methods

We calculated the total energies and electronic structures of all the pnictides considered using spin-polarized DFT as implemented in the Vienna ab initio simulation package [17, 18] and the projector augmented wave (PAW) potentials [19], with the specific PAW potentials used compiled in the electronic supporting information (ESI). We used a plane wave basis set with an energy cutoff of 520 eV, as benchmarked by our previous study on pnictides as PV candidates [4]. In all structural relaxation calculations, we allowed the cell shape, cell volume, and atomic positions to change without preserving any underlying symmetry until the total energies and atomic forces converged to below $10^{-5}$ eV and |0.01| eV/Å, respectively, for all pnictides considered. To choose the optimal Γ-centred *k*-point density for sampling the irreducible Brillouin zone, we calculated the total energies of $CaSrZnP_2$ with 32 and 48 *k*-points per Å and found negligible difference (< 0.1 meV/atom) in the calculated energies between the two *k*-point densities. Hence, we used a *k*-point density of 32 *k*-points per Å for all structural relaxations in this work to reduce computational costs. We used a Gaussian smearing width of 0.05 eV to integrate the Fermi surface and treated the exchange-correlation interactions with the strongly constrained and appropriately normed (SCAN [20]) functional for all the structural relaxation and the electronic DOS calculations.

To decide the compositions of pnictides for screening, we queried the popular materials databases, namely, the inorganic crystal structure database (ICSD [21]), the materials project [22], the open quantum materials database [23] and the nomad repository [24] for pnictides with X = N, or P; A = Li, Na, or K; B = Mg, Ca, Sr, or Zn; C = Al, Ga, or In; and D = Si, Ge, Ti, Sn, or Pb. X, A, B, C, and D groups correspond to elements that typically exhibit a -3 (anion), +1, +2, +3, and +4 oxidation state, respectively. We chose the above list of elements as they are relatively earth-abundant, mostly non-toxic (except Ga and Pb), and redox-inactive (hence less prone to forming charge-compensated point defects). Based on allowed stoichiometries that maintain charge neutrality with these elements, we listed unique ternary and quaternary pnictide classes, such as ABX, $B_3XX'$, $D_3X_4$, $A_3X$, $CC'X_2$, $ABCX_2$, $BDX_2$, $A_2DX_2$, $A_4BX_2$ and $BB'B''X_2$ (here ' and " represent non-identical elements from the same elemental set). Subsequently, we queried for the permutations of all possible compositions of each class of pnictides considered in the available databases for their experimentally/computationally reported crystal structures, electronic band gaps and/or previous reports of being considered as alternative PV materials. Among the pnictide classes, we chose $ABCX_2$ (72 possible compositions), $A_4BX_2$ (24) and $BB'B''X_2$ (8) for this work since other pnictide classes were either well-studied before (for their electronic properties) and/or did not have any experimental/computational structures.

In order to initialize the structures of the $A_4BX_2$ class, we considered the experimentally characterized crystal structures of $K_4ZnP_2$ ($R\bar{3}mH$ space group, ICSD collection code 67261) and the primitive cell of $Li_4SrN_2$ ($I4_1/amd$, collection code 87413) as templates. Similarly, for the $ABCX_2$ class, we used the experimentally reported structures of $LiCaGaN_2$ ($P2_1/c$, collection code 424911) and $LiSrGaN_2$ ($C2/m$, collection code 96224) as templates. Since we did not find suitable experimentally observed structures for the BB'B"$X_2$ class, we chose previously reported theory-derived structures that were predicted to be thermodynamically stable as templates [4], namely, $CaSrZnN_2$ ($I4/mmm$) and $CaSrZnP_2$ ($P\bar{3}m1$). See **Figure 1** for the structures of the templates considered in this work.

Subsequently, we initialized the structure of each composition according to the two templates representing each pnictide class by replacing the atoms of the template with the corresponding atoms of like-oxidation states, i.e., we performed simple chemical substitutions to generate hypothetical structures of a target composition. The structure that yielded the lowest DFT total energy upon structure relaxation among the theoretical structures considered for a composition (i.e., the ground state polymorph) was chosen for subsequent DFT calculations for that composition. To evaluate the thermodynamic stability of all pnictide compositions considered, we constructed the 0 K phase diagrams (or convex hulls) using the 'PhaseDiagram' class of the pymatgen package [25]. For each pnictide composition, we considered the DFT calculated energies of all possible ordered elemental, binary, ternary, and quaternary structures, as available in the ICSD, within the corresponding chemical space (for example, the K-Sr-Zn-P chemical space was considered for $KSrZnP_2$).

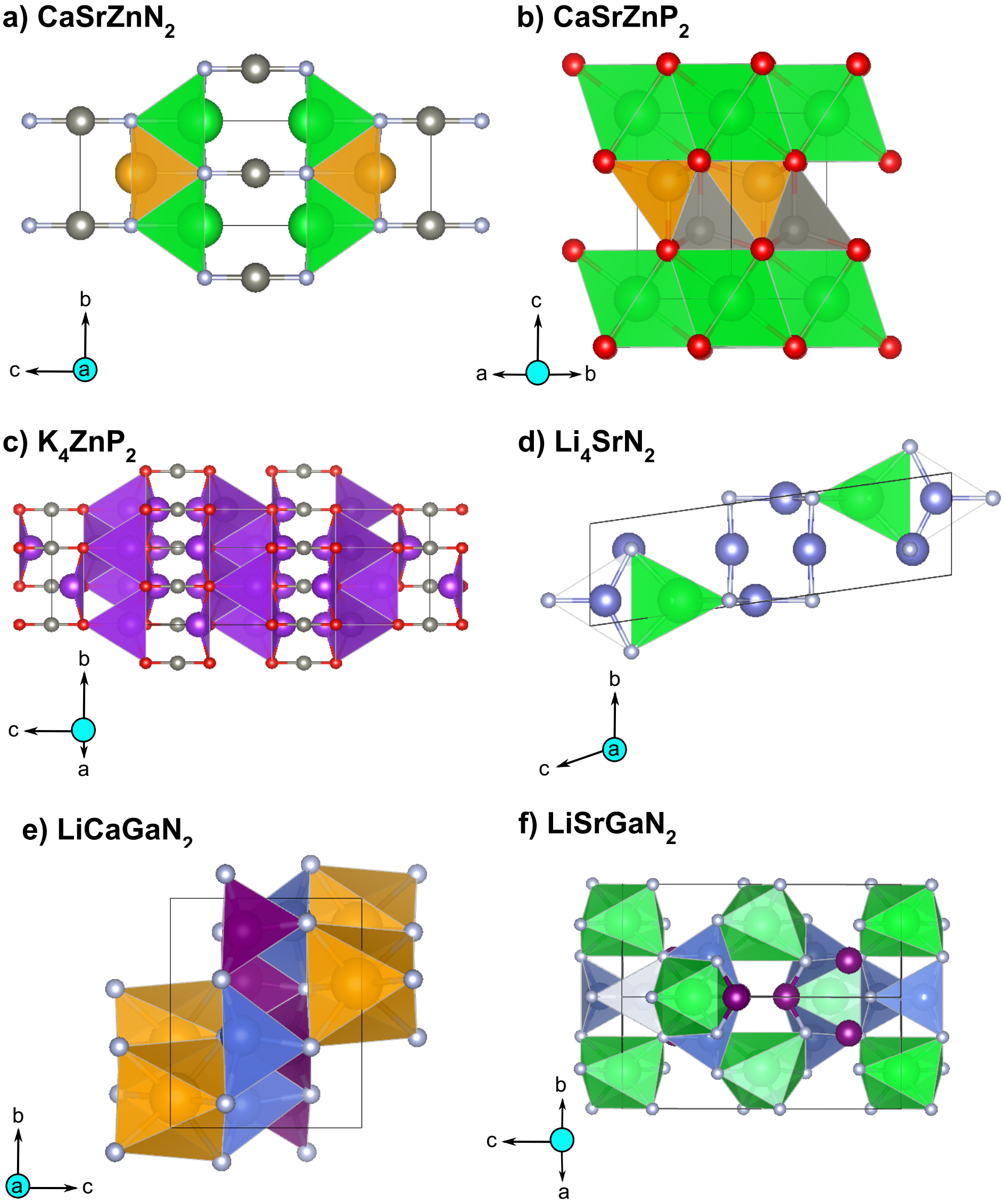

**Figure 1**: The template structures (a) $CaSrZnN_2$ and (b) $CaSrZnP_2$ for the BB'B''$X_2$ class, (c) $K_4ZnP_2$ and (d) the primitive cell of $Li_4SrN_2$ for the $A_4BX_2$ class, and (e) $LiCaGaN_2$ and (f) $LiSrGaN_2$ for the $ABCX_2$ class. The spheres with different colours denote different atom types: light purple for K, grey for Zn, red for P, dark blue for Li, light blue for N, green for Sr, white for N, yellow for Ca, and dark violet for Ga.

For a rapid screening of band gaps of the stable (or metastable) pnictides from each composition class, we performed the electronic DOS calculations with a self-consistent field (SCF) energy cutoff of $10^{-6}$ eV, a *k*-mesh density of 96 *k*-points per Å, and the improved linear tetrahedron method [26] for smearing. For the candidates with a suitable SCAN-calculated band gap, we performed further band structure calculations using the Heyd-Scuseria-Ernzerhof (HSE06[27]) hybrid functional with an SCF energy cutoff of $10^{-6}$ eV, since SCAN typically underestimates band gaps of semiconductors [3, 4]. For the HSE06 band structure calculations, we identified the high-symmetry *k*-points for the primitive cell of the structures considered according to [28]. Subsequently, we added zero-weighted $k$-points between the high-symmetry $k$-points for the band structure calculation, as implemented by the vaspkit software [29, 30].

For a select set of candidate pnictides, we calculated the effective masses of the electrons and holes that influence carrier mobility. Specifically, we calculated the effective masses at the $k$-point that shows the smallest energy of transition from the valence band to the conduction band, i.e., the $k$-point with the nearest-direct band gap, in the HSE band structure calculation. Note that for direct band gap semiconductors, this choice of the $k$-point is trivial. Mathematically, the electron and hole effective masses are given as, $m_e^* = \frac{\hbar^2}{\left|\left(\frac{\partial^2 E_c}{\partial k^2}\right)\right|}$ and $m_h^* = \frac{-\hbar^2}{\left|\left(\frac{\partial^2 E_v}{\partial k^2}\right)\right|}$, respectively, where $E_c$ and $E_v$ are the energies of the conduction and valence bands, and $\hbar$ is the modified Planck's constant. Using the scipy package [31], we performed cubic-spline interpolation of the $E$-values of the conduction and valence bands with the corresponding $k$-values, to numerically calculate the curvature of $E$ vs. $k$, using the method of central differences [32] as,

$$\frac{\partial^2 E_i}{\partial k^2} \approx \frac{E_{i+1} - 2E_i + E_{i-1}}{(k_i - k_{i-1})^2}$$

where $E_i$ is the energy of the $i^{\text{th}}$ interpolated point, and $k_i$ is the corresponding value of $k$. The value of $i$ is chosen such that $k_i$ represents the (nearest) direct band gap of the material.

We used the phonopy package [33] to perform phonon DOS calculations with the SCAN functional to determine the thermal properties and dynamic stability of candidate pnictides. The real-space force constants were calculated using $\pm 0.01$ Å symmetrically unique displacements of all atoms in a given unit cell. For each displacement, we performed an SCF with an energy convergence criterion of $10^{-6}$ eV, sampled at the $\Gamma$ point. The thermal properties and phonon DOS were evaluated on a 32×32×32 mesh, and the data is presented in **Figures S4** and **S5** of the ESI. The thermal properties were calculated excluding the imaginary phonon modes that are indicated by the negative frequencies in **Figure S5**.

For the set of four final candidates identified in our screening process, namely $K_4ZnP_2$, $Na_4ZnP_2$, $NaCaInN_2$ and $NaSrInN_2$, we calculated the intrinsic point defect formation energies of cation and anion single vacancies and cation anti-site pairs to determine their resistance to the formation of such defects. We used 2×2×2 supercells of $NaCaInN_2$ and $NaSrInN_2$, and the 2×2×1 supercells of $K_4ZnP_2$ and $Na_4ZnP_2$ to model the defective structures to ensure minimal interactions between the defects and their periodic images. The formation energy of any neutral defect ($E^f$) is, as given by [34],

$$E^f = E_{defect} - E_{bulk} - \sum_i n_i \mu_i$$

Where $E_{defect}$ and $E_{bulk}$ are the total energies of the defective and pristine supercells, respectively, $n_i$ is the number of $i$-atoms removed (negative) or added (positive) to form the defective configuration, with $\mu_i$ being its corresponding chemical potential, as calculated from the 0 K convex hull. The ranges of $\mu$ used for the defects considered are listed in **Table S3** of ESI.

# 3 Results

## 3.1 Structures and thermodynamic stability

While **Figure 1** shows the unit cells and atomic positions of the template structures used for generating the initial structures for the BB'B"$X_2$, $A_4BX_2$ and $ABCX_2$ pnictide classes, **Figure 2** compiles the energy above the convex hull ($E^{Hull}$), as heat maps, for all the ground state polymorphs of all pnictide compositions considered. Blue (red) squares indicate structures that are stable/metastable (unstable) with the corresponding $E^{Hull}$ (in meV/atom) denoted as numerical annotations. We use a $E^{Hull}$ threshold of 30 meV/atom for a compound to be considered as metastable and hence potentially synthesizable, as indicated by the solid green lines in **Figure 2**, based on prior work [4, 16, 35]. Although a previous study [35] has shown that pnictides having $E^{Hull}$ values below 100 meV/atom can be metastable, we impose a stricter $E^{Hull}$ threshold since compounds with higher $E^{Hull}$ values will be prone to form point defects.

The presence of an asterisk (*) in a given square in **Figure 2** indicates the corresponding ground state polymorph of that composition. For example, in **Figure 2a**, compositions with the asterisk have the $I4/mmm$-derived structure (from $CaSrZnN_2$, **Figure 1a**) being the ground state while those without the asterisk have the $P\bar{3}m1$-derived structure (from $CaSrZnP_2$, **Figure 1b**) as the ground state. Similarly, in **Figure 2b**, the asterisk indicates that the $I4_1/amd$-derived structure (from $Li_4SrN_2$, **Figure 1d**) to be the ground state, while in panels c-f, the asterisks signify the ground states to be $C2/m$-derived structures (from $LiSrGaN_2$, **Figure 1f**).

‘nan’ squares in **Figure 2a** represent compositions that are non-applicable/repeated for the BB’B”$X_2$ class.

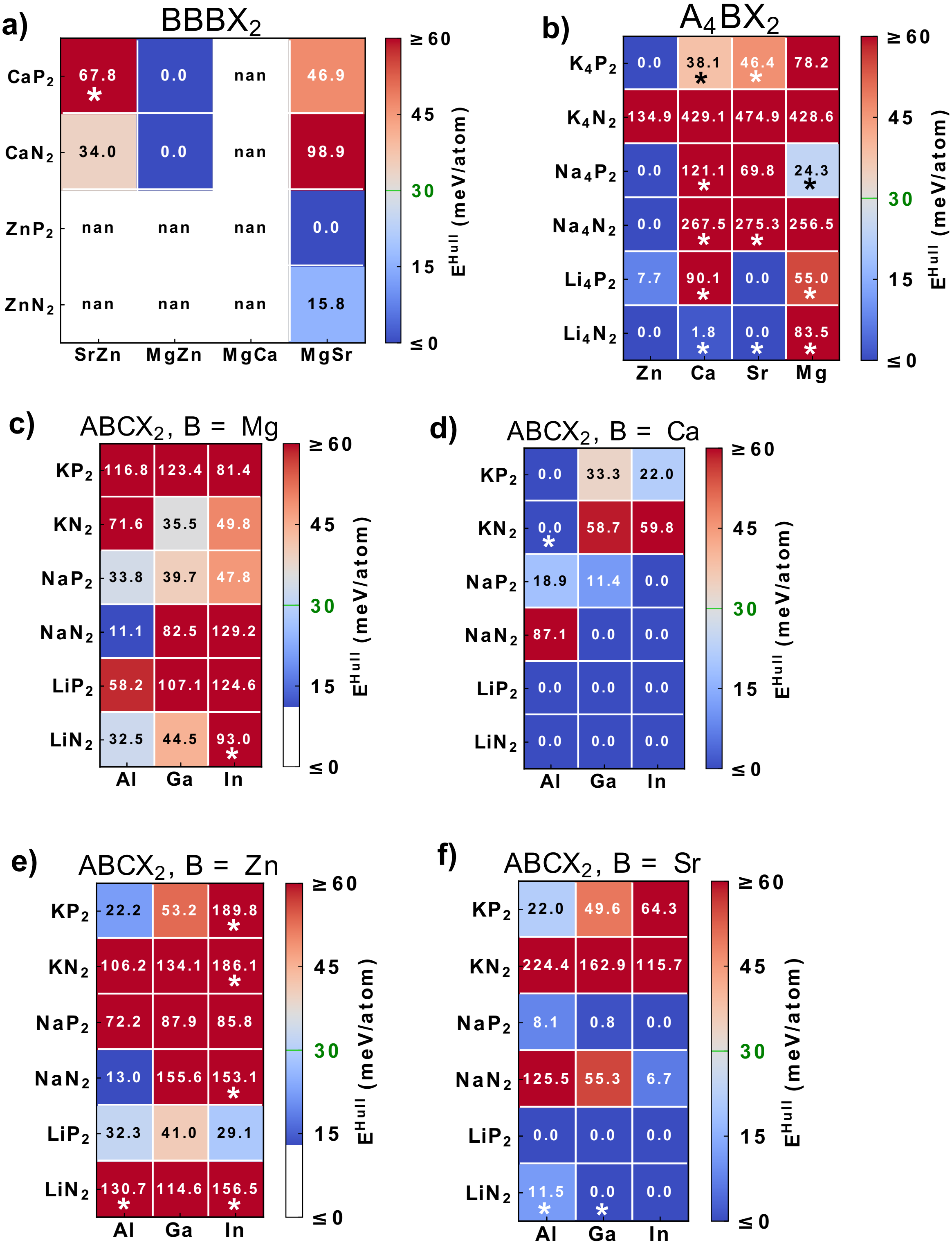

**Figure 2**: Heatmaps showing the calculated $E^{Hull}$ (in meV/atom, denoted as numbers within each square) for compounds of the (a) BB’B”$X_2$ class (denoted as $BBBX_2$), (b) $A_4BX_2$ class, and $ABCX_2$ class with B = Mg (c), Ca (d), Zn (e), and Sr (f). ‘nan’ denotes non-applicable or repeated compositions of the BB’B”$X_2$ class. Each row and column in a panel represent the combination of elements to form the composition within a given square.

In the $A_4BX_2$ class (**Figure 2b**), 11 compositions prefer the $I4_1/amd$-derived structure as their ground state configuration, while 13 favour the $R\bar{3}mH$ structure, indicating similar preferences for both templates. Li-compounds frequently adopt the $I4_1/amd$ structure (5/8 instances), likely due to $Li_4SrN_2$ being the template (**Figure 1d**), with the tendency of the $I4_1/amd$ being the ground state decreasing with increasing A cation size (4/8 instances with Na and 2/8 instances with K). Moreover, the $R\bar{3}mH$ template (**Figure 1c**) has the A cations being tetrahedrally coordinated, which may be better suitable for large cations such as Na and K. Within the $ABCX_2$ class (panels c-f in **Figure 2**), only nine of 72 compositions prefer the $C2/m$-derived structure, with Zn-containing compositions accounting for five out of the nine compositions, all of which are unstable. With respect to BB'B"$X_2$ compounds (**Figure 2a**), our findings of the preferred polymorphs of $CaSrZnN_2$ and $CaSrZnP_2$ align with those obtained from the previous study [4], while the remaining compositions prefer the $P\bar{3}m1$ template.

In terms of $E^{Hull}$, we find only 41 out of the 104 possible compositions to exhibit values that are $\leq$30 meV/atom threshold, and can be considered as metastable and potentially synthesizeable. Specifically, we find 26 of the 41 compositions predicted to be thermodynamically stable (i.e., $E^{Hull}$ = 0 meV/atom), with the distribution of three compositions in the BB'B"$X_2$ class, six compositions in $A_4BX_2$ class, and the remaining 17 compositions in $ABCX_2$ class, which is dominated by 11 compositions in Ca-based pnictides (**Figure 2d**). The choice of the template structure may have played a significant role in identifying thermodynamically stable structures for a given pnictide composition, given that Ca-based pnictides predominantly prefer the Ca-containing $LiCaGaN_2$ structure as the template (**Figure 1e**) and are thermodynamically stable as well. Similarly, several Mg based compositions are predicted to be unstable, which may be due to the non-availability of Mg-containing structures as templates. Thus, we do highlight the possibility of previously unexplored (or unnoticed) structures that can be suitable templates for compositions currently classified as unstable (see Discussion section).

## 3.2 Electronic structure

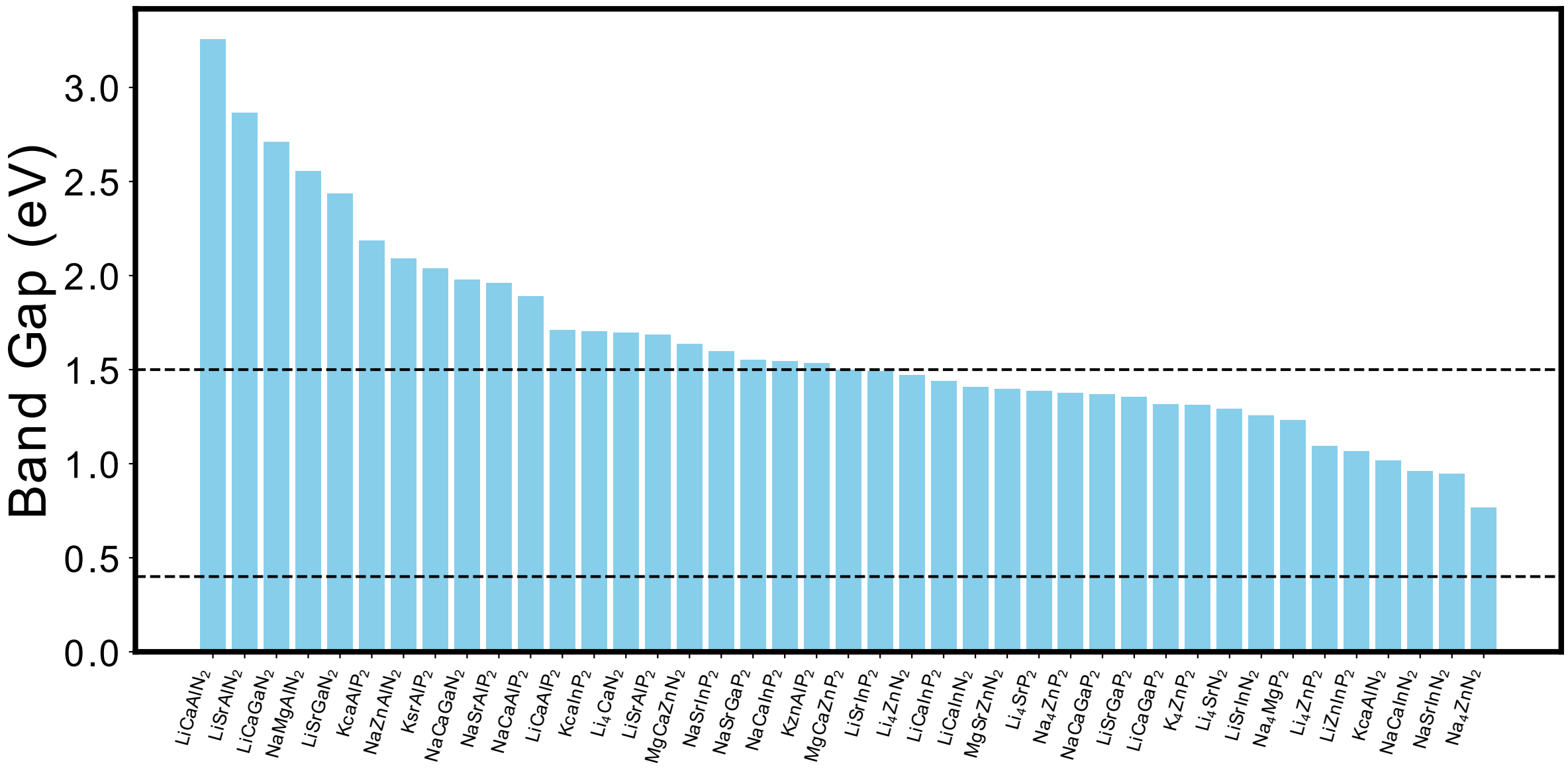

**Figure 3**: SCAN-calculated band gaps obtained from the electronic DOS calculations on the thermodynamically stable and metastable pnictides.

For the 41 stable and metastable candidates, we perform electronic DOS calculations to screen for optimal band gaps based on SCAN calculations, as shown in **Figure 3**. Each bar in **Figure 3** represents a pnictide compound, with the dashed horizontal lines showing the range of SCAN-calculated band gaps (i.e., 0.4 eV to 1.5 eV) that we consider for subsequent screening. Note that our previous work [4] has shown that compared to HSE06, SCAN underpredicts the band gaps in pnictides by ~0.7 eV. Hence, for the purpose of swift screening on band gaps with SCAN that minimizes the number of false negative candidates, we reduce the Shockley-Queisser lower bound of 1.1 eV by 0.7 eV, hence setting a minimum SCAN-calculated band gap of 0.4 eV as a screening filter. Given that SCAN generally tends to underestimate the band gap for all semiconductors and insulators, we keep the upper limit of our screening to 1.5 eV, which is typically used as the upper bound of the Shockley-Queisser limit.

Some of the qualitative trends that can be observed from the data in **Figure 3** include the decreasing band gap among $ABCN_2$ compounds from C = Al to In for a given combination of A and B, while in the case of $ABCP_2$ compounds, the band gaps are lowest for A and B combinations with C = Ga. Further, **Figure 3** indicates the existence of 22 compositions that exhibit SCAN-calculated band gaps in the 0.4 to 1.5 eV range, for which we subsequently perform band structure calculations with the HSE06 hybrid functional to verify the magnitude and nature of the band gap. Of the 22 compounds, three belong to the BB'B"$X_2$ class, eight belong to the $A_4BX_2$ class, and the remaining 11 belong to the $ABCX_2$ class. Data from the HSE06 calculations are compiled in **Table S1** of the ESI. We also plot the trends in SCAN versus HSE06 band gaps among the 22 compounds, in the form of a parity plot in **Figure S1**, which highlights the underprediction of band gaps by SCAN compared to HSE06, particularly for nitrides, justifying the need for HSE06 based band structure estimates in addition to the SCAN-based swift screening. Note that the only exception in our pnictide space where the SCAN-calculated band gap is higher than HSE06 is for the $Na_4ZnN_2$ compound, which HSE06 predicts to be metallic.

Based on the HSE06 calculated band structures, we consider the compounds with a direct band gap or a nearest direct band gap (for indirect band gap compounds) in the 1.1 to 1.75 eV as candidate compositions for further calculations. In indirect band gap compounds, although the energy of light absorbed may be greater than the band gap value, the optical transitions at a conserved wave vector (i.e., a given *k*-point) are allowed. The minimum energy of such conserved wave vector transitions is what we refer to as the nearest direct band gap. To ensure that a significant part of the solar spectrum can be absorbed during these 'direct' transitions, we set an upper bound of 1.75 eV for the HSE06-calculated (nearest direct) band gap of a pnictide for the material to be considered a candidate.

**Figure 4** shows the HSE06 band structures of the five candidate pnictide compounds we identify for further calculations, namely, $NaCaInN_2$ (panel a), $NaSrInN_2$ (b), $K_4ZnP_2$ (c), $Na_4ZnP_2$ (d), and $Na_4MgP_2$ (e). **Figures S2** and **S3** of the ESI compile the HSE06 band structures of the remaining 17 non-candidate pnictides (see **Table S1** for the full list). The spin up and spin down states in **Figures 4**, **S2**, and **S3** are shown in solid blue and dashed orange curves, respectively, while the horizontal green lines indicate the valence band maximum (VBM) and conduction band minimum (CBM) with the zero of the energy scale referenced to the VBM. In the case of $Na_4ZnN_2$, the horizontal green line corresponds to the Fermi level (**Figure S2e**). Vertical dashed lines in all band structure plots indicate high-symmetry *k*-points, relevant for the corresponding crystal structure considered.

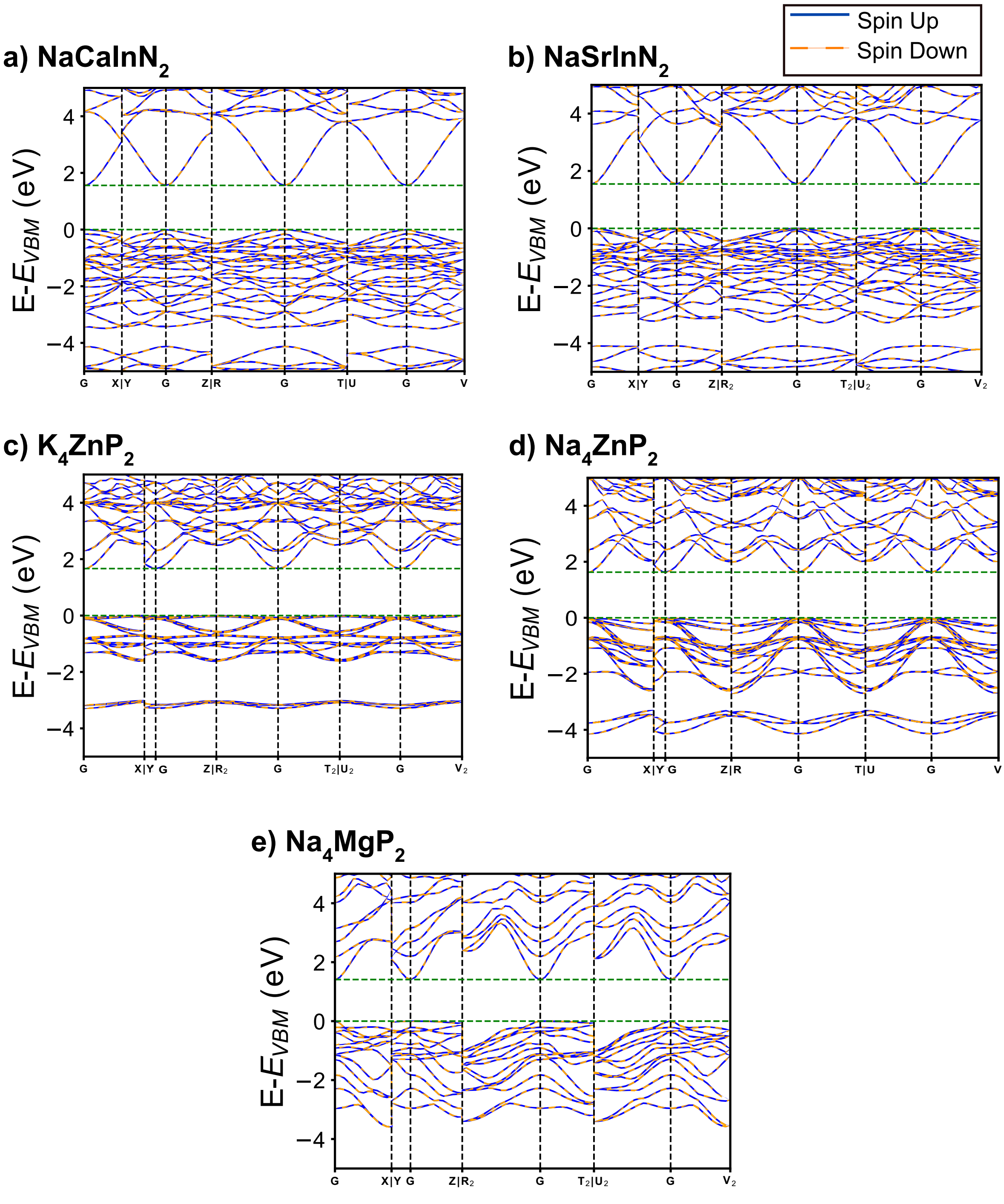

**Figure 4**: HSE06-calculated band structures of the candidate pnictide compounds: (a) $NaCaInN_2$, (b) $NaSrInN_2$, (c) $K_4ZnP_2$, (d) $Na_4ZnP_2$ and (e) $Na_4MgP_2$ using the HSE06 hybrid functional.

Among the candidates identified in **Figure 4**, $NaCaInN_2$ (panel a), $NaSrInN_2$ (b), and $Na_4MgP_2$ (e) exhibit direct band gaps at their corresponding Γ-points (denotes as 'G' in **Figures 4, S2,** and **S3**), with values of 1.56 eV, 1.55 eV, and 1.41 eV, respectively. $K_4ZnP_2$ (**Figure 4c**) and $Na_4ZnP_2$ (**Figure 4d**) are indirect band gap semiconductors and have their CBM at the Γ-point and their VBM at Z and Y, respectively. While the indirect band gaps of $K_4ZnP_2$ and $Na_4ZnP_2$ are 1.66 eV and 1.63 eV, respectively (**Table S1**), their nearest direct band gaps, occurring at the Γ-point for both compositions, are 1.73 eV and 1.66 eV, respectively, which are within our HSE06 screening range. Note that our thermodynamic stability calculations (**Figure 2**) indicate $NaCaInN_2$, $K_4ZnP_2$, and $Na_4ZnP_2$ to be thermodynamically stable ($E^{Hull}$=0 meV/atom), while $NaSrInN_2$ ($E^{Hull}$=6.7 meV/atom) and $Na_4MgP_2$ ($E^{Hull}$=24.3 meV/atom) to be metastable. Thus, we have three candidates in the $A_4BX_2$ class, two candidates from the $ABCX_2$ class, and none from the BB'B"$X_2$ class that filter through our thermodynamic stability and band gap screening. We consider the five pnictides identified in **Figure 4** for further evaluations on effective mass and dynamic stability (see below).

### 3.3 Effective masses

To validate the numerical procedure to calculate effective carrier masses, we benchmark our calculations on a well-known semiconductor, GaAs, and compare the results with a previous study [36], and tabulate the findings in **Table S2** of the ESI. Importantly, we observe a good agreement with our calculations and literature especially on the effective hole masses arising from the heavy valence band of GaAs and the effective electron mass. Hence, we use the same procedure (as done for GaAs) to estimate the hole and electron effective masses, $m_h^{eff}$ and $m_e^{eff}$, respectively, for the five pnictide candidates at their corresponding Γ points and tabulate them in **Table 1**. Note that we use our HSE06-calculated band structures for estimating the effective masses.

**Table 1**: Calculated effective masses of holes and electrons ($m_h^{eff}$ and $m_e^{eff}$) in units of free electron mass (~ $9.1\times10^{-31}$ kg) for the five pnictide candidates considered.

| Compound | $m_h^{eff}$ | $m_e^{eff}$ |
|---|---|---|
| $NaCaInN_2$ | -0.26 | 0.18 |
| $NaSrInN_2$ | -0.89 | 0.17 |
| $K_4ZnP_2$ | -39.32 | 0.28 |
| $Na_4ZnP_2$ | 0.46 | 0.21 |
| $Na_4MgP_2$ | -0.02 | 0.18 |

In general, compounds having carriers with small masses (i.e., negative and positive masses for holes and electrons, respectively) exhibit superior carrier mobilities and are therefore preferred for PV applications. Among the five candidates considered, $NaCaInN_2$ and $NaSrInN_2$ show lower $m_e^{eff}$ (~0.17-0.18, **Table 1**) compared to the corresponding $|m_h^{eff}|$ (~0.26-0.89), suggesting higher mobility of electrons than holes at the band extrema. This can be attributed to the greater curvature of the conduction band in $NaCaInN_2$ and $NaSrInN_2$ compared to the corresponding valence bands at the Γ point (panels a and b of **Figure 4**). In contrast, holes are the more mobile carriers with a lower $|m_h^{eff}|$ (~0.02) than $m_e^{eff}$ (~0.18) in $Na_4MgP_2$. For $K_4ZnP_2$ (**Figure 4c**), the valence band is nearly flat, leading to a large $|m_h^{eff}|$ (~39.32) at Γ and limiting its mobility, while the material should exhibit reasonable electron mobility ($m_e^{eff}$~0.28). In the case of $Na_4ZnP_2$ (**Figure 4d**), the valence band displays an upward curvature at Γ, reflecting in a positive $m_h^{eff}$ (~0.46) that could limit its hole mobility compared to its electron mobility ($m_e^{eff}$~0.21). Thus, off the five candidates considered, we do not envision limitations due to carrier mobility in $NaCaInN_2$, $NaSrInN_2$, and $Na_4MgP_2$, while doping $K_4ZnP_2$ and $Na_4ZnP_2$ may be a useful strategy to improve their $p$-type conductivity.

### 3.4 Dynamic stability and defect formation energies

To evaluate the dynamic stability of the candidate pnictides, we computed the phonon DOS and the corresponding thermal properties, as compiled in **Figures S4** and **S5**. For the calculation of the thermal properties in **Figure S5**, we did not consider any imaginary phonon modes in **Figure S4**. Given $Na_4MgP_2$ is prone to the formation of point defects (*vide infra*), we did not compute the phonon DOS for the structure. Importantly, we do not find any imaginary phonon modes for $NaCaInN_2$ (**Figure S4a**), while we observe a few imaginary modes for $NaSrInN_2$ (**Figure S4b**) and $K_4ZnP_2$ (**Figure S4c**), which can be attributed to numerical noise in our DFT calculations. Thus, we can conclude that the $NaCaInN_2$, $NaSrInN_2$, and $K_4ZnP_2$ structures are dynamically stable. On the other hand, we do observe a significant number of imaginary phonon modes for $Na_4ZnP_2$ (**Figure S4d**) highlighting its dynamical instability. Importantly, we observe $Na_4ZnP_2$ to be a dynamically unstable structure, even though it is predicted to be thermodynamically stable (**Figure 2**), which can be attributed to the limited choice of template structures explored, highlighting the need for the screening of dynamic stability in addition to thermodynamic stability as well. Thus, $Na_4ZnP_2$ may still be a candidate that can be explored as a PV candidate, if the precise ground state structure that is dynamically stable is identified. Overall, we consider $NaCaInN_2$, $NaSrInN_2$, $K_4ZnP_2$, and $Na_4MgP_2$ for performing defect calculations.

The energies associated with the formation of intrinsic point defects for the four pnictide compounds considered are shown in **Figure 5**. Specifically, we consider the formation of A, B, and C cation vacancies ($Vac_A$, $Vac_B$, and $Vac_C$, respectively, in **Figure 5**), the X anion vacancy ($Vac_X$), and the cation anti-site pairs ($A_B$+$B_A$, $B_C$+$C_B$ and $C_A$+$A_C$) for the $ABCX_2$ compounds, namely, $NaCaInN_2$ and $NaSrInN_2$. For the $A_4BX_2$ compounds, $Na_4MgP_2$ and $K_4ZnP_2$, we calculate the $E^f$ of A and B cation vacancies ($Vac_A$ and $Vac_B$), the X anion vacancy ($Vac_X$) and the cation anti-site pair ($A_B$+$B_A$). For example, we calculate the $E^f$ of $Vac_{Na}$, $Vac_{Ca}$, $Vac_{In}$, $Vac_N$, $Na_{Ca}$+$Ca_{Na}$ (i.e., Na on Ca site + Ca on Na site), $Ca_{In}$+$In_{Ca}$, and $Na_{In}$+$In_{Na}$ defects in $NaCaInN_2$, while we calculate the $E^f$ of $Vac_K$, $Vac_{Zn}$, $Vac_P$, and $K_{Zn}$+$Zn_K$ defects in $K_4ZnP_2$. While the solid bars for the vacancy defects in **Figure 5** represent the lowest $E^f$ for each defect, the shaded regions represent the variation in the vacancy $E^f$ with changes in the $\mu$ of the species removed (see **Table S3**). The dashed horizontal line in **Figure 5** represents a threshold of 1 eV used for identifying candidates that are susceptible to the formation of point defects. Despite being an arbitrary value, our previous study [4] used 1 eV as a threshold since point defects with $E^f$ above 1 eV have shown minimal impact on the PV performance of other candidate materials, such as kesterite-based chalcogenides [37, 38, 39].

Importantly, all candidates except $Na_4MgP_2$, i.e., $NaCaInN_2$, $NaSrInN_2$, and $K_4ZnP_2$, exhibit $E^f$ for several defects that are generally greater than the 1 eV threshold. All pnictides consider do exhibit $Vac_X$ $E^f$<1 eV, especially in the cases of $NaCaInN_2$ and $NaSrInN_2$, indicating their poor resistance to $n$-type defects such as $Vac_X$. While the $Vac_X$ is the only defect that falls below the 1 eV threshold in $K_4ZnP_2$, $NaCaInN_2$ ($Na_{Ca}$+$Ca_{In}$) and $NaSrInN_2$ ($Na_{Sr}$+$Sr_{In}$ and $Na_{In}$+$In_{Na}$) do exhibit additional cation anti-site pairs that lie below the threshold. Nonetheless, given that $NaCaInN_2$ and $NaSrInN_2$ exhibit $E^f$>1 eV for the other point defects considered, their predicted thermodynamic (meta)stability (**Figure 2**), and their suitable direct band gaps (**Figure 4**), we recommend them as novel candidates to be considered for PV applications. Notably, $K_4ZnP_2$ is an experimentally characterized compound that has been studied for its optoelectronic properties [13], and our calculations do indicate that its synthesis in a P-rich environment could avoid $Vac_P$ formation (hashed grey bar in **Figure 5**). This highlights the promise of $K_4ZnP_2$ for PV applications with some required doping to improve its hole mobility (**Table 1**). In the case of $Na_4MgP_2$, the lowest $E^f$ of all defects considered are significantly <1 eV, with the $Na_{Mg}$+$Mg_{Na}$ anti-site pair being predicted to form spontaneously ($E^f$< 0 eV). Note that $Na_4MgP_2$ is predicted to be thermodynamically metastable ($E^{Hull}$~24.3 meV/atom), which can be a contributory factor to its tendency to form point defects. Thus, given the propensity of $Na_4MgP_2$ to form point defects, we do not expect it to be reliable PV candidate. Overall, we propose $NaCaInN_2$, $NaSrInN_2$, and $K_4ZnP_2$ as the promising pnictide materials to be considered for PV applications.

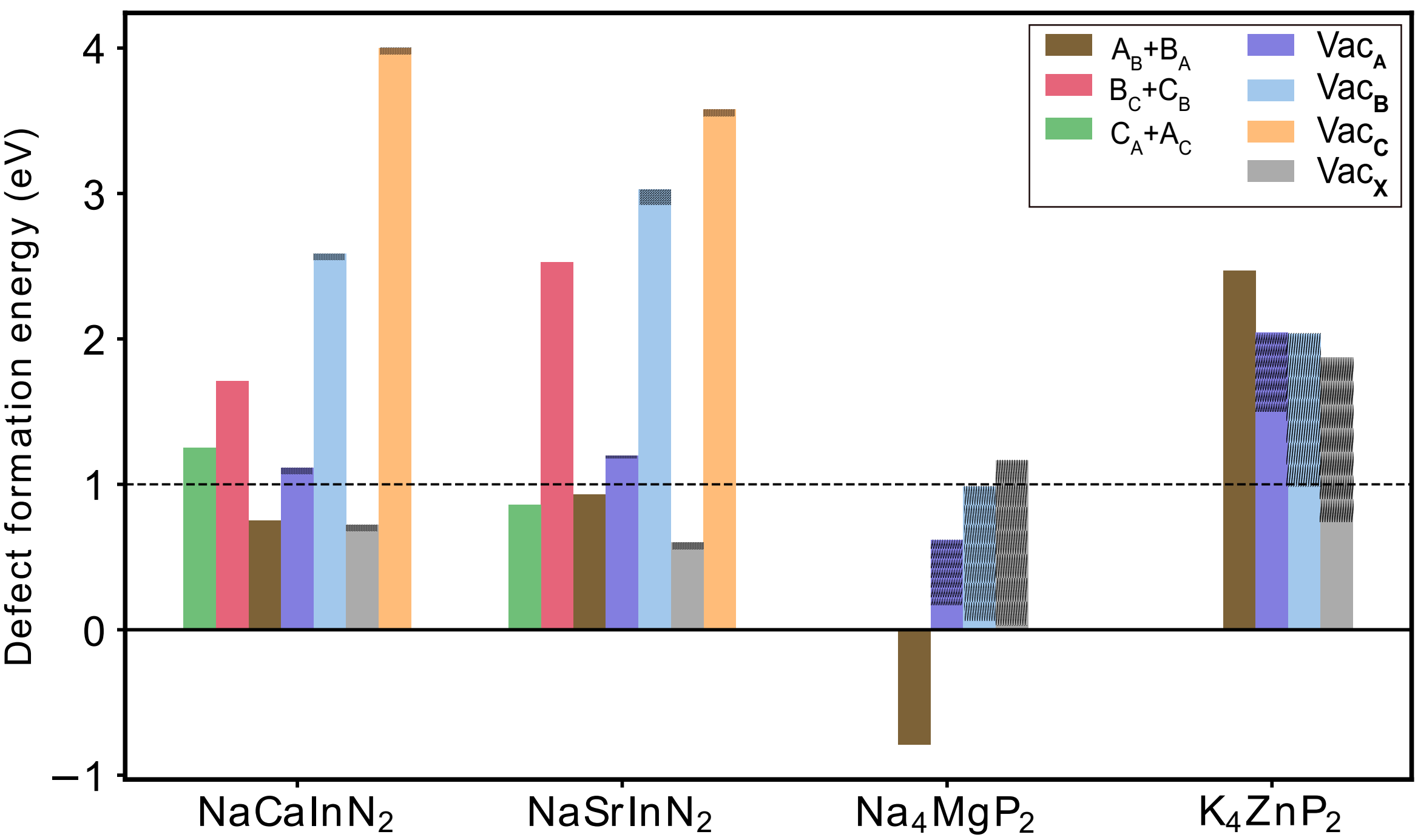

**Figure 5**: Point defect formation energies of the cation vacancies of A ($Vac_A$, violet), B ($Vac_B$, light blue), C ($Vac_C$, yellow) atoms, vacancy of the X anion ($Vac_X$, grey) and the cation anti-site pairs, namely, $A_B$+$B_A$ (brown), $B_C$+$C_B$ (pink) and $C_A$+$A_C$ (green) for the candidate pnictides considered. Due to the absence of the C atom in $Na_4MgP_2$ and $K_4ZnP_2$, the bars corresponding to $Vac_C$, $B_C$+$C_B$, and $C_A$+$A_C$ defects are not shown.

## 4 Discussion

In this study, we have explored three classes of ternary and quaternary pnictides comprising of monovalent, divalent and trivalent metal-based nitrides and phosphides as possible PV materials using DFT-based calculations. Considering a set of 104 pnictide compositions of three classes, BB'B"$X_2$, $A_4BX_2$, and $ABCX_2$, we initialized the pnictide structures based on known structures as templates (**Figure 1**), assessed the 0 K thermodynamic stability of the stable polymorph (**Figure 2**), and filtered the (meta)stable compounds based on their electronic structure via band gaps calculated with SCAN (**Figure 3**) and band structures and carrier effective masses calculated using HSE06 (**Figure 4** and **Table 1**). Finally, we evaluated the dynamic stability (**Figure S4**) and the resistance to the formation of point defects (**Figure 5**) for a subset of pnictide candidates, resulting in the identification of $NaCaInN_2$, $NaSrInN_2$, and $K_4ZnP_2$ as prospective PV candidates. Thus, our study represents a collection of systematic filters and calculated data that can be used in identifying novel PV materials in previously unexplored chemical spaces.

Throughout our pnictide screening, we have not considered binary pnictide compositions explicitly as possible candidates since pre-existing literature has explored them computationally or experimentally [4, 11, 40, 41]. However, we have considered binary pnictides for the calculation of the 0 K convex hulls and the associated estimation of the $E^{Hull}$ to identify (meta)stable candidates (**Figure 2**). Another chemical space that we did not explore in our work explicitly is ternary/quaternary pnictides based on group-14 elements (such as Si) that are redox-inactive, non-toxic, and abundant, due to the lack of reliable structures that can be used as templates. Note that the advent of generative models for materials can enable the exploration of chemical spaces for which reasonable structures are not known *a priori*, but we did not use such models in our work since we wanted to keep the structure generation part rooted in established literature.

One possible limitation of our structure generation process is the consideration of only two possible structural templates for each pnictide class (**Figure 1**), which does not preclude the existence of structures that are (more) stable both thermodynamically and dynamically. For instance, a better choice of a structural template can possibly yield a dynamically stable structure for $Na_4ZnP_2$ (**Figure S4d**). However, there exists pnictide compositions with the same structures used in our work but not considered by us as pnictide candidates due to the toxicity of Cd, such as $K_4CdP_2$ and $Na_4CdP_2$. Moreover, a recent study has characterized the $MgCaZnN_2$ composition to exhibit the $P\bar{3}m1$ space group [42], similar to what we have obtained as the stable polymorph from our workflow, thus reinforcing the reliability of our workflow.

We have used the SCAN-calculated band gaps as a relatively inexpensive filter to screen pnictides (**Figure 3**) before subjecting them to more accurate HSE06-based band structure calculations (**Figure 4**). Additionally, we have used different ranges of optimal band gap values for our SCAN and HSE06 based screenings, since SCAN is known to systematically underestimate band gaps compared to HSE06 (**Figure S1**). Although HSE06 is known to predict band gaps with reasonable accuracy, it does not guarantee the best theoretical predictions [4, 38, 43], which usually require computationally expensive quasi-particle calculations. Nevertheless, our workflow combining both SCAN and HSE06 calculations should reduce the number of false negative candidates that we may have missed in our screening.

In terms of point defects, we have accounted for vacancies and cation anti-site pairs in our work (**Figure 5**). However, we have not evaluated the resistance of the candidates towards forming other defects, such as interstitial, extrinsic, and line defects, which can also contribute to the eventual performance of the material as a PV. Notably, the band structure of a compound is subject to change when the compound is doped (with $n$- or $p$-type dopants) or is made part of interfaces (such as a heterojunctions) to form an efficient PV device. The structural and electronic properties also vary with the dimensionality of the structure, such as 2-D sheets and 1-D nanoribbons [44, 45]. The variations in the structural and electronic properties of the pnictides with topology and/or dopants require further detailed set of calculations, which are beyond the scope of this work.

Since the defect calculations are done at 0 K, we have ignored the entropic stabilization of defective configurations, which can be non-negligible at non-zero temperatures [46, 47]. Nevertheless, we expect our data to be useful in guiding experiments and identifying the important defects to optimize (such as $Vac_X$, **Figure 5**) within the candidates. Additionally, the synthesis of $NaSrInN_2$, one of the candidates identified in our work, may require higher temperatures and/or pressures, since we predict the structure to be metastable ($E^{Hull}$~6.7 meV/atom) and the formation of $Vac_N$ also needs to be minimized. Also, pnictides are usually sensitive to air and moisture, that can result in the formation of ammonia (for nitrides) or the toxic phosphine (for phosphides) during synthesis, operation or degradation [4, 48-50], highlighting the need to estimate and/or manage the moisture stability of the pnictide candidates.

## 5 Conclusion

The development of novel semiconductors consisting of non-toxic and earth-abundant elements that can be efficient PVs is required for transitioning fully to fossil-free and non-depleting sources of energy. Thus, we have comprehensively and systematically investigated ternary and quaternary pnictides, using DFT-based calculations, as candidate materials for PV applications. Specifically, we examined compositions of $A_4BX_2$, $BB'B''X_2$, and $ABCX_2$ pnictide classes (A = Li, Na, or K; B, B', B'' = Ca, Sr, Mg, or Zn; C = Al, Ga, or In; X = N, P) resulting in a total of 104 possible pnictide compositions and determined their ground state structure and thermodynamic stability using DFT. Subsequently, for a subset of screened candidates, we evaluated their electronic structure (using both SCAN and HSE06 functionals), carrier effective masses, dynamic stability, and their resistance to the formation of intrinsic point defects. Our screening approach resulted in a set of three promising candidates, namely, $NaCaInN_2$, $NaSrInN_2$, and $K_4ZnP_2$. Among the candidates, we predict $NaCaInN_2$ and $K_4ZnP_2$ to be thermodynamically stable ($NaSrInN_2$ is metastable), while $NaCaInN_2$ and $NaSrInN_2$ exhibit direct band gaps ($K_4ZnP_2$ exhibits an indirect band gap). Notably, our initial screening of thermodynamic stability and electronic structure did identify two additional candidates, namely, $Na_4ZnP_2$ which is dynamically unstable, and $Na_4MgP_2$ which is prone to the formation of cation anti-site pair defects, highlighting the need for including multiple calculable properties as screening criteria to identify reliable candidates. We hope that our study will reinvigorate attention in the pnictide chemical space for further exploration and provide useful guidance in the synthesis, optimization, and characterization for PV applications and beyond.

**Electronic Supporting Information**

PAW potentials used, HSE06 band gap data, SCAN-HSE06 band gap comparison, benchmarking effective masses in GaAs, and the range of chemical potentials used for defect calculations on all structures considered.

**Data availability**

All the computational data presented in this study are freely available to all on our GitHub repository, https://github.com/sai-mat-group/pnictides-pv-screening.

**Acknowledgements**

G.S.G. acknowledges financial support from the Indian Institute of Science. A.B. thanks the Indian Institute of Science Education and Research, Pune for financial support. The authors gratefully acknowledge the computational resources of the super computer 'Param Pravega' provided by Super Computer Education and Research Centre (SERC), IISc, where some of the density functional theory calculations showcased in this work were done.